\documentclass{article}

\usepackage{graphicx}

\begin{document}

%%\begin{flushright}
%%Liverpool Preprint: LTH \\
%%\end{flushright}

%%\draft  % \draft command makes pacs numbers print

\vspace{20mm}
\begin{center}
{\LARGE \bf The decay constant of the 
first excited pion from lattice QCD}\\[15mm] 

{\large\it UKQCD Collaboration}\\[3mm]

{\bf C. McNeile \\
Department of Physics and Astronomy, The Kelvin Building
University of Glasgow, Glasgow G12 8QQ, U.K.
}\\[2mm]
\begin{center} and  \end{center} 
{\bf C. Michael \\
Theoretical Physics Division, Dept. of Mathematical Sciences, 
University of Liverpool, Liverpool L69 3BX, UK 
}\\[2mm]

\end{center}

\begin{abstract}
We review the theory that predicts that
the decay constant of the excited light pseudo-scalar 
mesons are suppressed in the chiral limit relative 
to the pion decay constant. We compute the decay
constant of the first excited pion ($\pi^\prime$) 
using unquenched QCD
at a fixed lattice spacing with sea quarks of mass
as low as a third the mass of the strange quark.
The final result is very sensitive
to the improvement conditions. We obtain
$f_{\pi^\prime} / f_{\pi}$ = 0.078(93) in the
chiral limit.
\end{abstract}

\section{Introduction}

%%
%% lattice QCD and excited states
%%

Holl et al.~\cite{Holl:2004fr}  
have argued from Schwinger-Dyson equations and 
on general grounds
that the decay constants of the excited light $0^{-+}$ mesons should be 
suppressed relative to the pion decay constant. 
The decay constant of the first excited pion has also been
found to be small 
in~\cite{Dominguez:1977np,Andrianov:1989pi,Volkov:1996br,Elias:1997ya,Andrianov:1998kj,Maltman:2001gc,Diehl:2001xe,Lucha:2006rq}.
In table~\ref{tab:excitedPIONresults} we summarise the values for
the decay constant of the first excited pion from various models.
The predictions from these models for $f_{\pi^\prime}$ are 
indeed remarkably small. When we first heard about the result that the 
decay constant of the excited light pseudo-scalar could be
as small a tenth of the pion decay constant, our first reaction
was a combination of "that is remarkable" and "unbelievable".
In this letter we check these predictions using lattice QCD.

Experimentally, the second lightest meson with $0^{-+}$ quantum
numbers, made out of light quarks, is the $\pi(1300)$. In our lattice
calculation we use the notation $\pi^\prime$ for the first excited
light pseudo-scalar meson.

From a (naive) quark model perspective it is surprising for the
$\pi^\prime$ to have a highly suppressed decay constant. We define
highly suppressed to be that the decay constant is less than 10 \% of
the pion decay constant.  In the naive quark model the leptonic decay
constant for S-wave states is proportional to the wave-function at the
origin. The wave function of the excited $0^{-+}$ state would be
smaller than the ground state, because the wave-function is more
spread out than for the ground state. However, there is no obvious
reason for a dramatic suppression of the decay constant.  It is
instructive to consider the decay constants of mesons with heavy
quarks.  Lakhina and Swanson~\cite{Lakhina:2006vg} have recently
computed the leptonic decay constant of mesons in the charmonium and
bottonium systems and compared their values against experiment and
lattice gauge theory. For example, Lakhina and
Swanson~\cite{Lakhina:2006vg} get good agreement from their model and
the experimental results for the leptonic decay constants of the
$J/\psi$ ($411 \pm 7$ MeV) and $\psi'$ ($279 \pm 8$ MeV) mesons.  The
same decay constants have also been computed by
Dudek et al.~\cite{Dudek:2006ej} from quenched lattice QCD.  This type
of modest suppression of the value for the decay constant of the excited 
meson relative to the decay constant of the ground state
agrees with our (perhaps simple minded) intuition

Below is "simplistic"  theoretical argument for the 
suppression of the decay constant of excited light pseudo-scalars
Consider the axial current applied to the meson $X$ (here applied to 
mesons with $J^{PC} = 0^{-+}$).

The decay constant ($f_X$) of the state $X$ is defined by
\begin{equation}
\langle 0 \mid A_\mu \mid X \rangle = i f_X p_\mu
\end{equation}
where $A_\mu$ is the axial current. The partial conservation of the 
axial current~\cite{Jansen:1995ck}
is
\begin{equation}
\partial_\mu A_\mu = m_q \pi
\label{eq:partialAA}
\end{equation}
where $\pi$ is the interpolating operator for
pion states (pseudo-scalar density) and $m_q$ is the quark mass. 
Equation~\ref{eq:partialAA} is an operator relation,
hence is true between any states.
This allows us to write:
\begin{equation}
f_X = \frac{1} {m_X^2}  m_q \langle 0 \mid  \pi \mid X \rangle 
\label{eq:smallPION}
\end{equation}
For the lightest pseudo-scalar 
\begin{equation}
m_X^2 \propto m_q
\end{equation}
so $f_\pi$ is non-zero in the chiral limit. For a meson that
is not a Goldstone state, $m_X$ is non-zero in the 
chiral limit, hence $f_X$ will vanish in the chiral limit.

In the past, corrections to equation~\ref{eq:partialAA}
from excited  pion states~\cite{Dominguez:1977np,PhysRev166_1826}
have been considered for corrections to the Goldberger-Trieman 
relation. For example~\cite{Dominguez:1977np} introduced
extended PCAC 
\begin{equation}
\partial^\mu  A_\mu = \sum_{n} m_{\pi_n}^2 f_{\pi_n} \pi_n
\end{equation}
where $\pi_n$ is the interpolating
operator for the n-th excited light $0^{-+}$ meson.

The PDG~\cite{Eidelman:2004wy} 
quotes the mass of the $\pi(1300)$ as $1300 \pm 100$ MeV
with a decay width of between 200 to 600 MeV. The predominant decay
mode is to $\pi \pi \pi$ (this includes $\rho \pi$).
There is readable discussion about the experimental issues with the $\pi(1300)$ 
state in~\cite{Barnes:1996ff}.
Diehl and Hiller~\cite{Diehl:2001xe} estimate the leptonic decay
constant of the $\pi(1300)$ to be less than 8.4 MeV
from the experimental 
bounds on the decay 
of the $\tau$ to $\pi(1300) + \nu_\tau$
from CLEO~\cite{Eidelman:2004wy,Asner:1999kj}.
%%%
%%% Diehl normalisation, f_pi 131 MeV
%%%

The mass of the $\pi(1300)$ meson has been studied before using
lattice QCD. Some quenched studies show reasonable agreement with
experiment~\cite{Burch:2006dg,Yamazaki:2001er}. The $\pi(1300)$ decays
via the strong interaction, so quenched calculations would not be
expected to be  so accurate for this state.  Using the maximum
entropy method the CP-PACS collaboration obtained the mass of the
first excited pion to be 660(590) MeV in the continuum limit of
quenched QCD~\cite{Yamazaki:2001er}.  The CP-PACS
collaboration~\cite{Yamazaki:2001er} also computed the leptonic decay
constant for the first excited $\rho$ meson. The ratio of the leptonic
decay constant of the excited to ground rho meson, obtained by
CP-PACS~\cite{Yamazaki:2001er} was 0.41.  This is modest in accord
with our quark model intuition.  The unquenched lattice QCD
calculation by the MILC collaboration, using improved staggered
quarks, obtained the first excited pseudo-scalar to be at the mass
1362(41)(205) MeV~\cite{Aubin:2004wf}.

\begin{table}[tb]
\begin{center}
\begin{tabular}{|c|c|} \hline
Group & $f_{\pi^\prime}$ MeV   \\ \hline
Volkov and Weiss~\cite{Volkov:1996br}  & 0.68 \\  
Elias et al.~\cite{Elias:1997ya}  & $4.2 \pm 2.4$ \\ 
Maltman and Kambor~\cite{Maltman:2001gc} & $3.11 \pm 0.65 $ \\ 
Andrianov et al.~\cite{Andrianov:1998kj} &  $0.52 - 2.26$ \\ 
Kataev et al.~\cite{Krasnikov:1981vw,Kataev:1982xu}       & $4.3$    \\ \hline
%%% draper_magic  \vsbtc
\end{tabular}
  \caption{
Summary of the values of the $\pi^\prime$ decay constant
determined from models and sum rules. Our normalisation
convention is $f_\pi$ = 131 MeV.
}
\end{center}
\label{tab:excitedPIONresults}
\end{table}

\section{The lattice QCD calculation}

In~\cite{Allton:2004qq} the UKQCD collaboration reported results for
the pion decay constant from a two flavour unquenched calculation.
The masses of the first excited light pseudo-scalar meson were reported 
but not the respective decay constants. The paper by UKQCD used a variational
technique, where a matrix of correlators is fitted to a transfer 
matrix based model. This is one of the more reliable techniques 
to study excited states on the lattice. Variational and other lattice methods 
to extract properties of excited hadrons 
are reviewed in~\cite{McNeile:2003dy}.

In this paper we report the decay constant of the first excited light
pseudo-scalar meson. The calculation used two flavours of degenerate
sea quarks. The non-perturbatively improved clover action was
used for the quarks with the Wilson gauge action. The lattice volume
was $16^3\;32$ and $\beta = 5.2$. The lattice spacing is roughly 0.1 fm.
The data for the pion decay constant and the mass of the
$\pi^\prime$ state have already been published 
in~\cite{Allton:2004qq,Allton:2001sk}, so
we do not repeat that here.
Note that UKQCD also showed good
agreement between the pion decay constant computed by UKQCD and the
JLQCD collaborations~\cite{Aoki:2002uc}.

\begin{figure}
\begin{center}
\includegraphics[scale=0.3,angle=270]{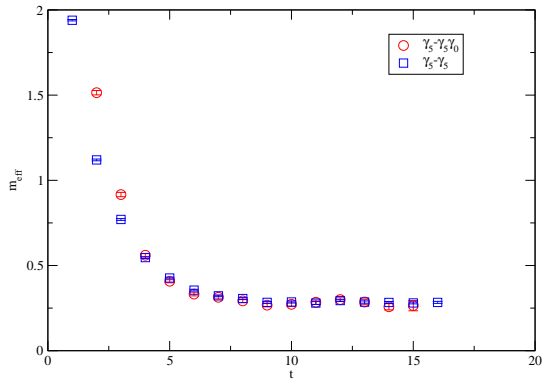}
\end{center}
\caption {
Effective mass for the $\gamma_5$ to $\gamma_5$ and for 
$\gamma_0 \gamma_5$ to $\gamma_5$ 
correlators for $\kappa$ = 0.1355.
 %%%XYZ  change fig inset to g5-g0g5
}
\label{fig:meff}
\end{figure}

One consequence of the suppression of the decay constants of the excited
pions would be that the correlators for $\gamma_0 \gamma_5$ 
to $\gamma_5$ should show less excited contamination  than for
the $\gamma_5$ to $\gamma_5$ correlator. In figure~\ref{fig:meff},
the effective mass plots for the two local correlators look
very similar. Recall, that the flatter the effective mass plot,
the less excited state contamination there is.

We use variational smearing to extract the 
required matrix elements. A 4 by 4 matrix of correlators was 
computed using as basis states: local operator, fuzzed operator
combined with the gamma matrix combinations $\gamma_5$ and 
$\gamma_0 \gamma_5$. Two mass states were included in the fits. 
The heaviest state included in the fit is likely to have more
excited state contamination in it.

%%
%% non-perurbative matching
%%
The lattice axial current requires both matching 
factors to convert to $\overline{MS}$ scheme, as well
as improvement factors. We use the formulation of the 
ALPHA collaboration~\cite{Jansen:1995ck}.
\begin{equation}
(A_R)^a_\mu = Z_A ( 1 + b_A m_q ) (A_I)^a_\mu
\end{equation}
We use the
non-perturbative renormalisation and improvement factors recently
computed by the ALPHA
collaboration~\cite{DellaMorte:2005se,DellaMorte:2005kg} for two
flavour unquenched QCD.  At $\beta = 5.2$ the ALPHA expressions give
$c_A =
-.0641$~\cite{DellaMorte:2005se}. At present, the $b_A$ factor is only
known to one loop in perturbation theory.

The $f_{\pi^\prime}$ decay constant is extracted from the 
following combination.
%%%%
\begin{equation}
f_{\pi^\prime} = Z_A  (1 + b_A m ) \left(a f_{A^\prime} + c_A f_{P^\prime} \right)
\label{eq:pionexciteRENORM}
\end{equation}
 %%%XYZ I think factors of m' are missing here
%%
where $f_{A^\prime}  = \frac{1}{m^\prime} \langle 0 \mid  \overline{\psi} \gamma_0 \gamma_5 \psi \mid \pi^\prime\rangle $ and 
$f_{P^\prime}  = \frac{1}{m^\prime} \langle 0 \mid \partial_4 \overline{\psi} \gamma_5 \psi \mid \pi^\prime\rangle $.

\begin{table}[tb]
\begin{center}
\begin{tabular}{|c|c|c|} \hline
$\kappa$ & a$f_{A^\prime}$  & a$f_{P^\prime}$ \\ \hline
0.1358   & $0.042_{+6}^{-8}$   & $0.585_{-7}^{+4}$  \\
0.1355   & $0.065^{+12}_{-8}$  & $0.60^{+6}_{-6}$   \\
0.1350   & $0.11^{+1}_{-2}$    & $0.73^{+6}_{-9}$   \\ \hline
%%% draper_magic  \vsbtc
\end{tabular}
  \caption{
Components of the $\pi^\prime$ decay constant. The ground state
decay constants are in the second table of~\cite{Allton:2004qq}.
The improved decay constant of the $\pi^\prime$  state is given
by equation~\ref{eq:pionexciteRENORM}.
}
\end{center}
\label{tab:excitedPIONOURresults}
\end{table}

\begin{figure}
\begin{center}
\includegraphics[scale=0.3,angle=270]{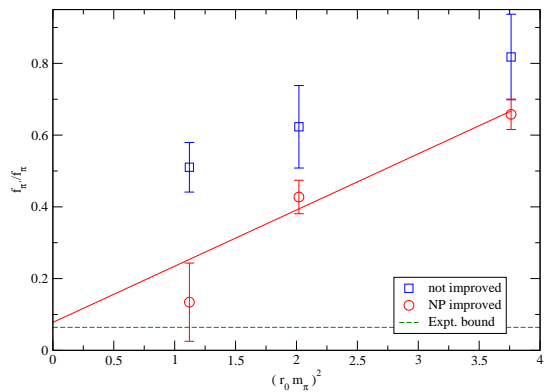}
\end{center}
\caption {Ratio of the decay constants of the first excited to ground
state light pseudo-scalar meson as a function of the pion mass squared.
The horizontal line is the experimental upper limit from
Diehl and Hiller~\cite{Diehl:2001xe}.
}
\label{fig:ratio}
\end{figure}

In figure~\ref{fig:ratio} we plot the ratio of the ratio of the decay
constants of the $\pi$ and $\pi'$ meson, using the unimproved and
improved decay constants from the ALPHA formulation. 
On the x-axis we use the square of the pion mass in 
units of $r_0$. The value of $r_0$ is not needed
for our final result, but a value of $r_0$ around 0.5 fm with
10\% errors can be used if required~\cite{Sommer:1993ce}.
The unimproved decay constant has only
a modest suppression of $f_{\pi^\prime}$ relative to $f_{\pi}$. The
value of $f_{\pi^\prime}$ obtained from the improved ALPHA formulation
is very much suppressed relative to $f_{\pi}$. In the continuum limit
the improved and un-improved decay constants should agree. However, at
the fixed lattice spacing that we use here, the improved decay
constant should be close to the continuum limit value of the decay 
constant because the $O(a)$ errors have been removed.

A simple linear fit to the ratio of decay constants 
with improvement gives $f_{\pi^\prime} / f_{\pi}$ = 0.078(93) in the 
chiral limit. The data at $\kappa=0.1358$ is important to
in producing a smaller extrapolated
ratio of decay constants.
A simple linear
extrapolation of the two heavier points gives $f_{\pi^\prime} / f_{\pi}$
= 0.16. For the unimproved decay constants, we obtain
$f_{\pi^\prime} / f_{\pi}$ = 0.38(11) in the 
chiral limit.
The modest suppression of the $f_{\pi^\prime}$ decay constant when no
improvement factors are used is consistent with the effective plots of
the axial to pseudo-scalar channel looking similar to the pseudo-scalar
to pseudo-scalar channel in figure~\ref{fig:meff}.  Although
figure~\ref{fig:ratio} could be viewed as a triumph of ALPHA's program
of non-perturbative renormalisation,
a controlled continuum extrapolation is required for a definitive
result.
Perhaps it is not so surprising 
that the full ALPHA formulation is required to see the suppression of the 
decay constant, because the PCAC relation is at the heart of the 
conditions imposed for improvement (determining coefficients of
irrelevant operators)~\cite{Jansen:1995ck}.
The increased suppression of the
excited pion decay constant, from the improved current,
is caused by the relatively large non-perturbative value of
the  coefficient of $c_A$ and the size of $f_{P^\prime}$.
In quenched QCD at $\beta=6.0$, with a similar lattice spacing 
to the one used here, the value of $c_A$ was controversial with 
different methods giving different 
results~\cite{Bhattacharya:2000pn,Collins:2001mm}. 
The ALPHA collaboration 
recommends that their formulation is used in a consistent continuum
extrapolation ("working at constant physics")~\cite{DellaMorte:2005se}. 
Also, there could
be excited state contamination in the coupling $f_{P^\prime}$, that
increases its value. To check the excited state contamination requires
a larger variational basis.

In the original paper the UKQCD collaboration noted
that the excited state masses were very close to
the mass of three pions~\cite{Allton:2004qq}.
It is not clear to us what we would expect for an axial
current insertion into a three pion state. If we
do a linear extrapolation of the masses of the
$\pi^\prime$ meson versus the square of the pion mass
in units of $r_0$ we obtain in the chiral limit:
$m_{\pi^\prime}$ = $1.29 \pm 0.29$ MeV (using a nominal
value of $r_0$ = 0.5 fm). Although the consistency 
of the extrapolated mass (within the large errors)
with the mass of the $\pi(1300)$
state is good, it will be more interesting in future
studies to monitor the effect of the three pion decay
on the state.

%%
%%  problems with the data set
%%

\section{Conclusion}

We have investigated the leptonic decay constant of the first excited
pseudo-scalar meson.
Although our results are consistent with $f_{\pi^\prime}$, being much
smaller than $f_{\pi}$, a definitive lattice QCD calculation requires
a continuum extrapolation. Recent algorithmic improvements have rescued
Wilson/clover lattice calculations
from being too computationally expensive 
with light quark masses,
hence it is now possible to repeat this calculation with more
than one lattice 
spacing and lighter 
quarks~\cite{Hasenbusch:2001ne,Urbach:2005ji,Luscher:2005mv}.

The suppression of the excited light pseudo-scalars may have 
implications for the proposed restoration of chiral symmetry for highly excited
hadrons, suggested by Glozman~\cite{Glozman:2003bt}. If the leptonic
decay constant of the excited pseudo-scalar mesons are suppressed, then
the leptonic decay constant of the  parity partners of the
pseudo-scalar mesons should also be suppressed.  
Although,
unfortunately, lattice QCD is not ideally suited to studying
highly excited mesons, the study of the suppression of the decay
constant of the first excited light pseudo-scalar is an interesting
first step. Note that even if lattice QCD cannot be used to study the 
proposed restoration of chiral symmetry in the excited hadron spectrum, it
may be able to study the proposed "restoration of chiral
symmetry" via the eigenvalue spectrum~\cite{DeGrand:2003sf,Cohen:2006bq}.

Following Narison~\cite{narison:2005wc}, 
in~\cite{McNeile:2006nv} we argued that for the
light $0^{++}$ state a small value for the leptonic decay constant
would be a signal for a molecular state. This seems an unlikely
interpretation for the smallness of the decay 
constant of the $\pi^\prime$ meson~\cite{Barnes:1996ff}.

The suppression of $f_{\pi^\prime}$ is a useful "benchmark" that can be 
used to tune and validate lattice QCD techniques that try to determine 
the properties of excited states mesons. It is particularly
interesting, because it is a consequence of chiral symmetry.
Holl et al.~\cite{Holl:2005vu} 
also have predictions for the sign of the 
decay constant of the $\pi^\prime$ meson that may also be studied
using lattice QCD.

\section{Acknowledgements}

We thank Craig Roberts for bringing this issue to our
attention. We thank Alexander Andrianov for comments.
This work has been supported in part by the EU Integrated
 Infrastructure Initiative Hadron Physics (I3HP) under contract
  RII3-CT-2004-506078. We are grateful to the ULgrid project of the
University of Liverpool for computer time. One of the authors (CM)
wishes to thank PPARC for the award of a Senior Fellowship.

%%\bibliographystyle{h-physrev2}
%%%\bibliography{q_mass} 

\begin{thebibliography}{10}

\bibitem{Holl:2004fr}
A.~Holl, A.~Krassnigg, and C.~D. Roberts,
\newblock Phys. Rev. {\bf C70}, 042203 (2004), nucl-th/0406030,
\newblock %%CITATION = NUCL-TH 0406030;%%.

\bibitem{Dominguez:1977np}
C.~A. Dominguez and M.~Moreno,
\newblock Phys. Rev. {\bf D16}, 856 (1977),
\newblock %%CITATION = PHRVA,D16,856;%%.

\bibitem{Andrianov:1989pi}
A.~A. Andrianov, V.~A. Andrianov, and A.~N. Manashov,
\newblock Int. J. Mod. Phys. {\bf A6}, 5435 (1991),
\newblock %%CITATION = IMPAE,A6,5435;%%.

\bibitem{Volkov:1996br}
M.~K. Volkov and C.~Weiss,
\newblock Phys. Rev. {\bf D56}, 221 (1997), hep-ph/9608347,
\newblock %%CITATION = HEP-PH 9608347;%%.

\bibitem{Elias:1997ya}
V.~Elias, A.~Fariborz, M.~A. Samuel, F.~Shi, and T.~G. Steele,
\newblock Phys. Lett. {\bf B412}, 131 (1997), hep-ph/9706472,
\newblock %%CITATION = HEP-PH 9706472;%%.

\bibitem{Andrianov:1998kj}
A.~A. Andrianov, D.~Espriu, and R.~Tarrach,
\newblock Nucl. Phys. {\bf B533}, 429 (1998), hep-ph/9803232,
\newblock %%CITATION = HEP-PH 9803232;%%.

\bibitem{Maltman:2001gc}
K.~Maltman and J.~Kambor,
\newblock Phys. Rev. {\bf D65}, 074013 (2002), hep-ph/0108227,
\newblock %%CITATION = HEP-PH 0108227;%%.

\bibitem{Diehl:2001xe}
M.~Diehl and G.~Hiller,
\newblock JHEP {\bf 06}, 067 (2001), hep-ph/0105194,
\newblock %%CITATION = HEP-PH 0105194;%%.

\bibitem{Lucha:2006rq}
W.~Lucha, D.~Melikhov, and S.~Simula,
\newblock (2006), hep-ph/0606281,
\newblock %%CITATION = HEP-PH 0606281;%%.

\bibitem{Lakhina:2006vg}
O.~Lakhina and E.~S. Swanson,
\newblock (2006), hep-ph/0603164,
\newblock %%CITATION = HEP-PH 0603164;%%.

\bibitem{Dudek:2006ej}
J.~J. Dudek, R.~G. Edwards, and D.~G. Richards,
\newblock Phys. Rev. {\bf D73}, 074507 (2006), hep-ph/0601137,
\newblock %%CITATION = HEP-PH 0601137;%%.

\bibitem{PhysRev166_1826}
C.~Michael,
\newblock Phys. Rev. {\bf 166}, 1826 (1968).

\bibitem{Eidelman:2004wy}
Particle Data Group, S.~Eidelman {\em et~al.},
\newblock Phys. Lett. {\bf B592}, 1 (2004),
\newblock %%CITATION = PHLTA,B592,1;%%.

\bibitem{Barnes:1996ff}
T.~Barnes, F.~E. Close, P.~R. Page, and E.~S. Swanson,
\newblock Phys. Rev. {\bf D55}, 4157 (1997), hep-ph/9609339,
\newblock %%CITATION = HEP-PH 9609339;%%.

\bibitem{Asner:1999kj}
CLEO, D.~M. Asner {\em et~al.},
\newblock Phys. Rev. {\bf D61}, 012002 (2000), hep-ex/9902022,
\newblock %%CITATION = HEP-EX 9902022;%%.

\bibitem{Burch:2006dg}
T.~Burch {\em et~al.},
\newblock Phys. Rev. {\bf D73}, 094505 (2006), hep-lat/0601026,
\newblock %%CITATION = HEP-LAT 0601026;%%.

\bibitem{Yamazaki:2001er}
CP-PACS, T.~Yamazaki {\em et~al.},
\newblock Phys. Rev. {\bf D65}, 014501 (2002), hep-lat/0105030,
\newblock %%CITATION = HEP-LAT 0105030;%%.

\bibitem{Aubin:2004wf}
C.~Aubin {\em et~al.},
\newblock Phys. Rev. {\bf D70}, 094505 (2004), hep-lat/0402030,
\newblock %%CITATION = HEP-LAT 0402030;%%.

\bibitem{Krasnikov:1981vw}
N.~V. Krasnikov and A.~A. Pivovarov,
\newblock Phys. Lett. {\bf B112}, 397 (1982),
\newblock %%CITATION = PHLTA,B112,397;%%.

\bibitem{Kataev:1982xu}
A.~L. Kataev, N.~V. Krasnikov, and A.~A. Pivovarov,
\newblock Phys. Lett. {\bf B123}, 93 (1983),
\newblock %%CITATION = PHLTA,B123,93;%%.

\bibitem{Allton:2004qq}
UKQCD, C.~R. Allton {\em et~al.},
\newblock (2004), hep-lat/0403007,
\newblock %%CITATION = HEP-LAT 0403007;%%.

\bibitem{McNeile:2003dy}
C.~McNeile,
\newblock (2003), hep-lat/0307027,
\newblock %%CITATION = HEP-LAT 0307027;%%.

\bibitem{Allton:2001sk}
UKQCD, C.~R. Allton {\em et~al.},
\newblock Phys. Rev. {\bf D65}, 054502 (2002), hep-lat/0107021,
\newblock %%CITATION = HEP-LAT 0107021;%%.

\bibitem{Aoki:2002uc}
JLQCD, S.~Aoki {\em et~al.},
\newblock Phys. Rev. {\bf D68}, 054502 (2003), hep-lat/0212039,
\newblock %%CITATION = HEP-LAT 0212039;%%.

\bibitem{Jansen:1995ck}
K.~Jansen {\em et~al.},
\newblock Phys. Lett. {\bf B372}, 275 (1996), hep-lat/9512009,
\newblock %%CITATION = HEP-LAT 9512009;%%.

\bibitem{DellaMorte:2005se}
M.~Della~Morte, R.~Hoffmann, and R.~Sommer,
\newblock JHEP {\bf 03}, 029 (2005), hep-lat/0503003,
\newblock %%CITATION = HEP-LAT 0503003;%%.

\bibitem{DellaMorte:2005kg}
ALPHA, M.~Della~Morte {\em et~al.},
\newblock Nucl. Phys. {\bf B729}, 117 (2005), hep-lat/0507035,
\newblock %%CITATION = HEP-LAT 0507035;%%.

\bibitem{Sommer:1993ce}
R.~Sommer,
\newblock Nucl. Phys. {\bf B411}, 839 (1994), hep-lat/9310022,
\newblock %%CITATION = HEP-LAT 9310022;%%.

\bibitem{Bhattacharya:2000pn}
T.~Bhattacharya, R.~Gupta, W.-J. Lee, and S.~R. Sharpe,
\newblock Phys. Rev. {\bf D63}, 074505 (2001), hep-lat/0009038,
\newblock %%CITATION = HEP-LAT 0009038;%%.

\bibitem{Collins:2001mm}
UKQCD, S.~Collins, C.~T.~H. Davies, G.~P. Lepage, and J.~Shigemitsu,
\newblock Phys. Rev. {\bf D67}, 014504 (2003), hep-lat/0110159,
\newblock %%CITATION = HEP-LAT 0110159;%%.

\bibitem{Hasenbusch:2001ne}
M.~Hasenbusch,
\newblock Phys. Lett. {\bf B519}, 177 (2001), hep-lat/0107019,
\newblock %%CITATION = HEP-LAT 0107019;%%.

\bibitem{Urbach:2005ji}
C.~Urbach, K.~Jansen, A.~Shindler, and U.~Wenger,
\newblock Comput. Phys. Commun. {\bf 174}, 87 (2006), hep-lat/0506011,
\newblock %%CITATION = HEP-LAT 0506011;%%.

\bibitem{Luscher:2005mv}
M.~Luscher,
\newblock PoS {\bf LAT2005}, 002 (2006), hep-lat/0509152,
\newblock %%CITATION = HEP-LAT 0509152;%%.

\bibitem{Glozman:2003bt}
L.~Y. Glozman,
\newblock Phys. Lett. {\bf B587}, 69 (2004), hep-ph/0312354,
\newblock %%CITATION = HEP-PH 0312354;%%.

\bibitem{DeGrand:2003sf}
T.~A. DeGrand,
\newblock Phys. Rev. {\bf D69}, 074024 (2004), hep-ph/0310303,
\newblock %%CITATION = HEP-PH 0310303;%%.

\bibitem{Cohen:2006bq}
T.~D. Cohen,
\newblock (2006), hep-ph/0605206,
\newblock %%CITATION = HEP-PH 0605206;%%.

\bibitem{narison:2005wc}
S.~Narison,
\newblock (2005), hep-ph/0512256,
\newblock %%citation = hep-ph 0512256;%%.

\bibitem{McNeile:2006nv}
UKQCD, C.~McNeile and C.~Michael,
\newblock (2006), hep-lat/0604009,
\newblock %%CITATION = HEP-LAT 0604009;%%.

\bibitem{Holl:2005vu}
A.~Holl, A.~Krassnigg, P.~Maris, C.~D. Roberts, and S.~V. Wright,
\newblock Phys. Rev. {\bf C71}, 065204 (2005), nucl-th/0503043,
\newblock %%CITATION = NUCL-TH 0503043;%%.

\end{thebibliography}

\end{document}